\pdfoutput=1

\documentclass[journal]{IEEEtran}
\ifCLASSINFOpdf
  \usepackage[pdftex]{graphicx}
  \DeclareGraphicsExtensions{.pdf,.jpeg,.png}
\else
  \usepackage[dvips]{graphicx}
  \DeclareGraphicsExtensions{.eps}
\fi
\usepackage{multirow}

\usepackage{array}
\newcolumntype{C}[1]{>{\centering\let\newline\\\arraybackslash\hspace{0pt}}m{#1}}

\begin{document}
%
\title{Acoustic scene classification using \\convolutional neural network and \\multiple-width frequency-delta data augmentation}
%
%
%

\author{Yoonchang~Han and~Kyogu~Lee,~\IEEEmembership{Senior Member,~IEEE}
\thanks{Y. Han and K. Lee are with the Music and Audio Research Group, Graduate School of Convergence Science and Technology, Seoul National University, Seoul 08826, Republic of Korea, e-mail: (yoonchanghan@snu.ac.kr, kglee@snu.ac.kr).}
\thanks{K. Lee is also with the Advanced Institutes of Convergence Technology, Suwon, Republic of Korea}
\thanks{Manuscript received \today; revised \today.}}

%
%

\markboth{Journal of \LaTeX\ Class Files,~Vol.~14, No.~8, August~2015}%
{Shell \MakeLowercase{\textit{et al.}}: Bare Demo of IEEEtran.cls for IEEE Journals}
%



\maketitle

\begin{abstract}
In recent years, neural network approaches have shown superior performance to conventional hand-made features in numerous application areas. In particular, convolutional neural networks (ConvNets) exploit spatially local correlations across input data to improve the performance of audio processing tasks, such as speech recognition, musical chord recognition, and onset detection. Here we apply ConvNet to acoustic scene classification, and show that the error rate can be further decreased by using delta features in the frequency domain. We propose a multiple-width frequency-delta (MWFD) data augmentation method that uses static mel-spectrogram and frequency-delta features as individual input examples. In addition, we describe a ConvNet output aggregation method designed for MWFD augmentation, folded mean aggregation, which combines output probabilities of static and MWFD features from the same analysis window using multiplication first, rather than taking an average of all output probabilities. We describe calculation results using the DCASE 2016 challenge dataset, which shows that ConvNet outperforms both of the baseline system with hand-crafted features and a deep neural network approach by around 7\%. The performance was further improved (by 5.7\%) using the MWFD augmentation together with folded mean aggregation. The system exhibited a classification accuracy of 0.831 when classifying 15 acoustic scenes.

\end{abstract}

\begin{IEEEkeywords}
DCASE 2016, acoustic scene classification, convolutional neural network, deep learning, multiple-width frequency-delta data augmentation.
\end{IEEEkeywords}

%
\IEEEpeerreviewmaketitle

\section{Introduction}
%
%
%
%
\IEEEPARstart{I}{n} the field of machine listening, recognizing environments has become a particularly important application. Acoustic scene classification (ASC) enables devices to make sense of their environment \cite{barchiesi2015acoustic}, and opens up a number of new applications. For instance, devices such as smart phones, internet-of-things (IoT) devices, wearable devices, and robots equipped with artificial intelligence, may all benefit from ASC by providing services and applications according to context.

In addition, intelligent personal assistants (IPAs) represent another field that can benefit from ASC. IPAs are software agents that make recommendations and perform actions automatically by analyzing various input data, including audio, images, user input, or contextual information such as location, weather, and personal schedules \cite{hauswald2015sirius}. IPA services such as Google's Google Now\footnote{http://www.google.com/landing/now/}, Microsoft's Cortana\footnote{https://www.microsoft.com/en/mobile/experiences/cortana/}, and Apple's Siri\footnote{https://www.apple.com/ios/siri/} make extensive use of audio input, and the use of contextual information extracted from environmental audio has significant potential for recommending appropriate actions to the user.

Sound event detection is a closely related research area to ASC. An acoustic scene may be thought as a collection of sound events on top of some ambient noise. For example, a `bus' scene may be identified from frequently occurring sound events such as acceleration, braking, passenger announcements, and door opening sounds, while the engine and other people's conversations exist in the background. Some approaches to ASC \cite{nam2013acoustic} exploit event detection techniques to increase the scene classification accuracy. ASC can also be used for improving sound detection performance \cite{heittola2013context}. ASC and sound event detection are closely related, and the boundary between them is often blurred \cite{barchiesi2015acoustic}. Here we limit the scope to identification of environmental sounds. 

There have been a number of approaches proposed for ASC over the past decades; however, there is a lack of common benchmarking datasets \cite{barchiesi2015acoustic}. The IEEE Audio and Acoustic Signal Processing (AASP) Technical Committee organized the first Detection and Classification of Acoustic Scenes and Events (DCASE) challenge in 2013, and then the DCASE 2016 challenge, with an extended ASC dataset. Over the past three years, a number of audio processing techniques have been proposed, and deep learning is arguably the most promising. As indicated by the term `deep learning', the method employs a high-level representation of low-level data by stacking multiple layers using nonlinear modules. There are several variants of deep learning architectures, and the convolutional neural network (ConvNet) method is a deep learning technique that is widely used for image classification, owing to its superior performance in learning distinctive local characteristics \cite{lecun2015deep}. 

There is growing interest in applying ConvNet for audio processing because the local characteristics of a time–-frequency representation of the audio signal contain important information on a number of classification tasks, as with computer vision. In the field of music information retrieval, approaches that utilize ConvNet can achieve state-of-the-art performance for tasks such as chord recognition \cite{humphrey2012rethinking}, onset detection \cite{schluter2014improved}, and music boundary detection \cite{ullrich2014boundary, grill2015music}. Furthermore, ConvNet has been applied to speech recognition \cite{abdel2014convolutional,abdel2013exploring,sainath2015deep}.

Here we investigate the ASC problem using ConvNet. The major contributions of this work are as follows.

\par \vspace{\baselineskip}
\hangindent=0.7cm 
1. We propose a ConvNet architecture for ASC with multiple-width frequency-delta (MWFD) data augmentation for input data arrangement in place of conventional channel stacking.
\par \vspace{\baselineskip}
\hangindent=0.7cm 
2. We propose a folded mean aggregation method for combining output probabilities from ConvNet with MWFD augmentation to classify audio clips at the scene level.
\par \vspace{\baselineskip}

The remainder of the paper is organized as follows. In Section II, we discuss various approaches to ASC. In Section III, the system architecture is described, including audio pre-processing and details of the proposed ConvNet architecture. The proposed MWFD data augmentation method and ConvNet output aggregation for audio clip-wise decision is also discussed. In Section IV, dataset specifications and the baseline (reference) system are examined, including the training configurations and the experimental settings used for evaluation. In Section V, the results are presented characterizing the performance of the proposed system for ASC, along with a comparison with existing algorithms. Here we focus on input data arrangement and aggregation and, in addition, investigate the intermediate outputs of the network to understand the behavior of the proposed model. We discuss the results of applying the proposed method to DCASE 2013 and 2016 datasets. Section VI concludes the paper.

\section{Background}
The DCASE 2013 challenge provides a good starting point for discussing research on topics such as ASC and sound event detection. It was composed of three tasks: ASC, event detection for synthetic data, and real-world audio. A total of 11 algorithms were submitted to the ASC aspect of this challenge, providing a useful benchmark for comparing the performance of algorithms. The benchmarking audio dataset included 10 scenes; i.e., a bus, a busy street, an office, an open-air market, a park, a quiet street, a restaurant, a supermarket, a metro train, and a metro train station. These datasets were recorded in the Greater London area. Each class included ten 30-s audio samples. Further details on these datasets can be found in \cite{giannoulis2013database}.

Most existing approaches are similar in that the system first extracts various audio features, and then makes decisions based on a classifier. Although the details of the feature extraction processes differ, manual extraction of audio features was the most popular method.

Nogueira et al. \cite{nogueira2013sound} used spectral features (i.e., mel-frequency cepstral coefficients (MFCCs)) along with temporal features, such as amplitude variation and standard deviation of MFCCs. Their approach also used spatial features such as interaural time/level differences and coherence of the two-channel stereo data. They used support vector machines (SVMs), and achieved an accuracy of 0.60. Chum et al. \cite{chum2013ieee} used spectral/temporal sparsity and loudness, with either SVMs or hidden Markov models (HMMs), and achieved an accuracy of 0.65.

Geiger et al. \cite{geiger2013recognising}, Patil and Elhilali \cite{patil2013multiresolution}, Elizalde et al. \cite{elizalde2013vector}, and Li et al. \cite{li2013auditory} used various combinations of spectral, temporal, energy, and voicing features with SVMs as a classifier, whereas \cite{elizalde2013vector} used Gaussian mixture models (GMMs) and \cite{li2013auditory} used TreeBagger for classification. The accuracy of these methods was in the range 0.58--0.72 and, in general, better performance was achieved than the baseline performance provided for the challenge (which corresponded to the use of MFCCs with a bag-of-frames approach, giving an accuracy of 0.55 \cite{giannoulis2013detection}).

On the other hand, some existing approaches have used features that are different from a typical hand-crafted features. Nam et al. \cite{nam2013acoustic} used sparse restricted Boltzmann machines (RBMs) to learn features from a mel-spectrogram. This feature learning approach achieved an accuracy of 0.60 using selective max-pooling and SVMs. Rakotomamonjy and Gasso \cite{rakotomamonjy2013histogram} used histograms of oriented gradient (HOG) features, extracted from a constant Q transform (CQT), which is widely used for object detection in image processing. The use of SVMs for classification achieved an accuracy of 0.69. Roma et al. \cite{roma2013recurrence} used recurrence quantification analysis (RQA) on features extracted from a similarity matrix computed using MFCCs, followed by SVMs for classification. They submitted two different settings for the challenge, and one of them achieved an accuracy of 0.76, which is the highest accuracy among all algorithms submitted to DCASE 2013 for the scene classification task.

As shown above, most of the previous approaches heavily rely on the manually designed audio features. However, acoustic scenes are highly complicated sound which contains various sound events and ambient sounds, and it is hard to design an audio feature that represents all important characteristics present in acoustic scenes. ConvNets exploit spatially localized correlations across input data to learn appropriate features. In the ASC task, this technique can be used for learning features that can describe unique sound events and ambient sounds in each acoustic scene. In the next section, we describe in detail the proposed system architecture using ConvNet.

\section{System Architecture}
\subsection{Audio Preprocessing}
ConvNet aims to learn high-level feature representation automatically from low-level data. Thus, appropriate pre-processing of input data is a crucial aspect of the system. In this section, we describe how we processed the input audio prior to feeding them into the ConvNet. First, we converted audio input to mono (from a stereo recording) by averaging the right and left channels, and normalized the data by dividing samples by the maximum absolute value, to restrict the amplitude range between -1 and 1. It is common to down-sample the audio to make the datasets smaller, as well as to remove inaudible frequencies; however, we used 44,100 Hz, as some of the scenes appeared to contain notable spectral characteristics at very high frequencies. Then, we performed the discrete Fourier transform (DFT) to obtain a time-frequency representation of the audio. For DFT, we used analysis frame size of 2,048 samples which is approximately 46-ms, with 50\% overlaps. Its linear frequency scale was then converted into a mel-frequency scale. A mel-scale is based on the human auditory system and is approximately logarithmic above 1 kHz \cite{logan2000mel}. We used 128 mel-frequency bins following representation learning researches on music annotation \cite{nam2012learning,hamel2011temporal}, musical instrument identification task \cite{han2016sparse}, and fingering detection of overblown flute sound \cite{han2016detecting}; this is a reasonable size that sufficiently retain the original spectral characteristics, while significantly reducing the dimensionality of the data.

The size of the analysis window is a critical factor affecting the classification accuracy. If the window is too small, it cannot contain sufficient information on the scene; if the window is too large, the temporal resolution will suffer and the number of examples becomes limited. To train the network, we used an analysis window length of 1-s, which has been reported to be optimal for similar ConvNet architecture \cite{2016arXiv160509507H}. Empirical experiments showed that the use of overlap did not improve performance, but did increase the computational complexity; therefore, we used non-overlapping windows so that the 30-s audio samples were divided into 30 analysis examples for training.

As a final step of preprocessing, we performed feature scaling by standardization. We subtracted the mean from the data and divided them with the standard deviation, hence the data have mean at zero with unit variance. We obtained the mean and standard deviation from the training data only, and testing data was standardized using the statistics obtained from the training data.

\begin{table}[!t]
\renewcommand{\arraystretch}{1.3}
\caption{Proposed ConvNet structure. The data shape indicates the number of filters $\times$ time $\times$ frequency. The activation function is followed by each convolutional layer, and then a fully connected layer.}
\centering
\begin{tabular}{l l}
    \hline
    \hline
    Data shape & Description\\
    \hline
    1 $\times$ 43 $\times$ 128 & mel-spectrogram\\
    1 $\times$ 45 $\times$ 130 & 1 $\times$ 1 zero-padding\\
    32 $\times$ 45 $\times$ 130 & 3 $\times$ 3 convolution, 32 filters\\
    32 $\times$ 47 $\times$ 132 & 1 $\times$ 1 zero-padding\\
    32 $\times$ 47 $\times$ 132 & 3 $\times$ 3 convolution, 32 filters\\
    32 $\times$ 15 $\times$ 44 & 3 $\times$ 3 max-pooling\\
    32 $\times$ 15 $\times$ 44 & dropout (0.25)\\
    32 $\times$ 17 $\times$ 46 & 1 $\times$ 1 zero-padding\\
    64 $\times$ 17 $\times$ 46 & 3 $\times$ 3 convolution, 64 filters\\
    64 $\times$ 19 $\times$ 48 & 1 $\times$ 1 zero-padding\\
    64 $\times$ 19 $\times$ 48 & 3 $\times$ 3 convolution, 64 filters\\
    64 $\times$ 6 $\times$ 16 & 3 $\times$ 3 max-pooling\\
    64 $\times$ 6 $\times$ 16 & dropout (0.25)\\
    64 $\times$ 8 $\times$ 18 & 1 $\times$ 1 zero-padding\\
    128 $\times$ 8 $\times$ 18 & 3 $\times$ 3 convolution, 128 filters\\
    128 $\times$ 10 $\times$ 20 & 1 $\times$ 1 zero-padding\\
    128 $\times$ 10 $\times$ 20 & 3 $\times$ 3 convolution, 128 filters\\
    128 $\times$ 3 $\times$ 6 & 3 $\times$ 3 max-pooling\\
    128 $\times$ 3 $\times$ 6 & dropout (0.25)\\
    128 $\times$ 5 $\times$ 8 & 1 $\times$ 1 zero-padding\\
    256 $\times$ 5 $\times$ 8 & 3 $\times$ 3 convolution, 256 filters\\
    256 $\times$ 7 $\times$ 10 & 1 $\times$ 1 zero-padding\\
    256 $\times$ 7 $\times$ 10 & 3 $\times$ 3 convolution, 256 filters\\
    256 $\times$ 1 $\times$ 1 & global average-pooling\\
    1024 & flattened and fully connected\\
    1024 & dropout (0.50)\\
    15 & softmax\\
    \hline
    \hline
\end {tabular}
\label{tab:convnet_arch}
\end{table}

\begin{figure*}[!t]
  \centering
  \includegraphics[keepaspectratio, width=\textwidth]{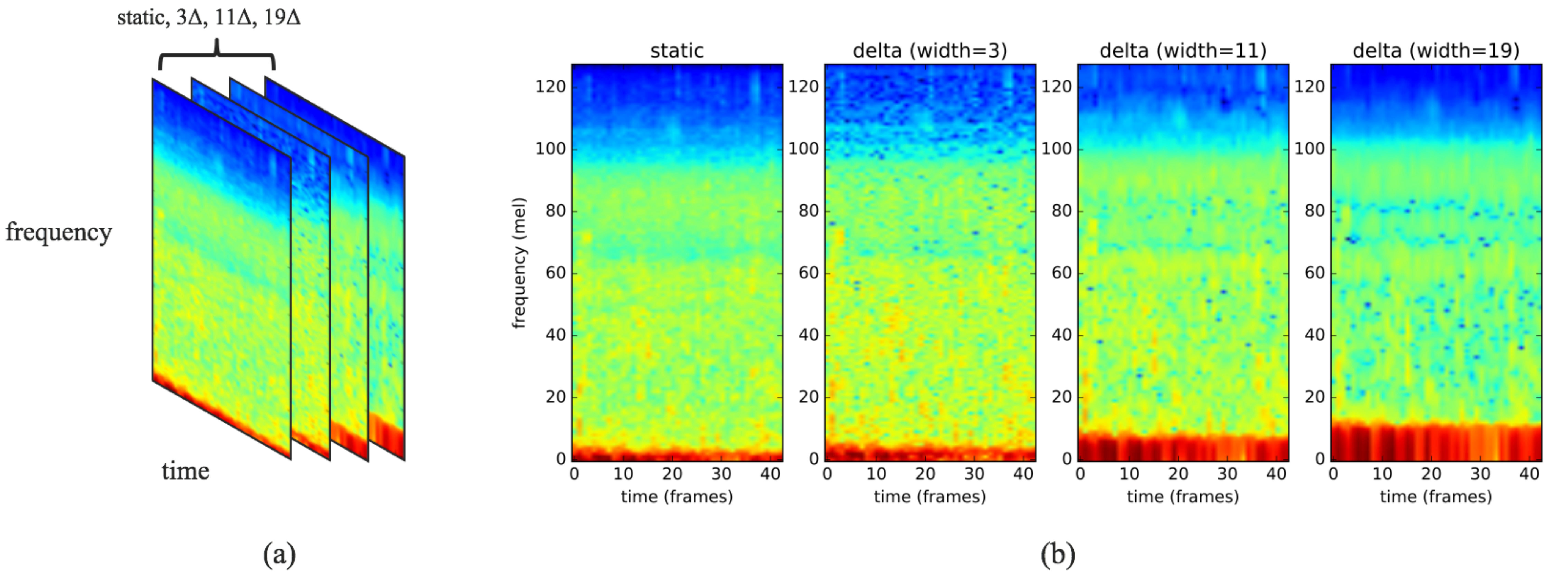}
  \caption{Two ConvNet input organization methods. (a) A typical method using several feature maps for ConvNet input. (b) Our MWFD data augmentation method, which uses frequency-delta features with different widths (here we used 3, 11, and 19), which are fed into ConvNet as individual examples together with static input with the same labels.}
\label{fig:figure_spread}
\end{figure*}

\subsection{Network Architecture}
Our ConvNet structure was inspired by VGGNet \cite{simonyan2014very}, which is composed of several layers of convolution, with a small receptive field (3$\times$3), followed by a max-pooling layer. VGGNet outperformed other ConvNet approaches for localization tasks at the ImageNet Challenge 2014, and this type of architecture has been adopted by numerous other researchers. We adopted the concept that stacking many layers of a small receptive field, but used less number of filters for each convolution layer (in the range 32--256) because the size of input data is relatively small. We added 1$\times$1 zero-padding prior to each convolution step to make full use of the data near the edges. The two consecutive convolution layers with a single max-pooling were termed `convolution block'. Four convolution blocks were used, incrementally increasing the number of filters, and included a fully connected layer at the end. The overall structure of the ConvNet algorithm is described in Table \ref{tab:convnet_arch}. It has been reported that global average pooling is a suitable replacement for fully connected layers to prevent over-fitting \cite{lin2013network}. However, we used a fully connected layer following the global average pooling layer, because our empirical experiments showed that this provided more stable performance. 

The choice of activation function is a critical factor affecting classification performance. The rectified linear unit (ReLU) was first introduced by Nair and Hinton \cite{nair2010rectified}, and is one of the most popular choices of activation function. Its non-saturating nonlinearity is particularly well suited to large-scale datasets, and enables faster learning than conventional saturating nonlinearities, such as sigmoid and hyperbolic tangent (tanh) \cite{krizhevsky2012imagenet}. It exhibits superior performance across various domains, and most recent studies using ConvNet have used this activation function \cite{simonyan2014very,li2015automatic,zeiler2014visualizing,sermanet2013overfeat}. It is defined as follows:

\begin{equation}
y_i = max(0,z_i)
\end{equation}

\noindent where $z_i$ is the input to the $i$th channel. Recently, several variations of ReLU have been proposed to further improve performance. Leaky ReLU is one such variant that was proposed by Mass et al. \cite{maas2013rectifier}. Leaky ReLU gives a small gradient in the negative part, whereas conventional ReLU suppresses the negative part to zero. This difference means that with leaky ReLU, some units may be active, which would have been inactive with normal ReLU. It is defined as follows:

\begin{equation}
y_i = \left\{
  \begin{array}{lr}
    z_i  &  z_i \ge 0\\
    \alpha_iz_i  &  z_i < 0
  \end{array}
\right.
\end{equation}

\noindent where $\alpha$ (which is valued between 0 and 1) describes the gradient of the negative part. It has been reported that Leaky ReLU outperforms ReLU for image classification \cite{xu2015empirical} and polyphonic instrument identification \cite{2016arXiv160509507H}. We used $\alpha = 0.33$, because it has been reported that a very leaky setting results in improved performance \cite{xu2015empirical}. The softmax function was used at the end of the network for classification.

Using the network structure and activation function described above, we achieved state-of-the-art performance in identification of the predominant instrument in real-world polyphonic music \cite{2016arXiv160509507H}. (Note that the network used in the proposed work uses global max-pooling instead of average pooling, and also uses softmax activation function instead of sigmoid function for classification compare to the previous work as it is a single-label problem.)

\subsection{Multiple-width Frequency-delta Data Augmentation}
In the field of image processing, it is common to handle the colors in the image by stacking them into several channels, each of which share the same local region \cite{schluter2014improved}; for example, using three channels for red, green, and blue (RGB). This de facto standard input data arrangement method is often similarly applied in audio processing. Grill and Schl\"uter \cite{grill2015music} decomposed input data using a harmonic-percussive source separation (HPSS) algorithm for two-channel data. Schl\"uter and B\"ock 
\cite{schluter2014improved} used three channels of spectrogram input data, each with different window sizes. Abdel-Hamid et al. \cite{abdel2014convolutional} used first and second temporal derivative (delta and double delta) features of mel-spectrogram for speech recognition.

In this section, we discuss how to organize time-frequency representations of audio data that are suitable for ConvNet. As mentioned above, typically there are several possible time-frequency representations of an audio signal that share the same locality (e.g., delta and double delta), which can be used for feature maps and fed in as multi-channel ConvNet input data. ConvNet uses a relatively small window, which moves across the input data, and each neuron in the convolution layer is connected only to a local region, but with a full depth (i.e., all channels). Using features containing different colors (as in computer vision), it appears likely that stacking them as several feature maps will be the optimal method of exploiting the available information. 

However, feature manipulation method such as delta is different from color map of the image. The original color of the image at a given location may only be described by the sum of several colors such as RGB, but delta features cannot be decomposed as part of the original version. Rather, delta features emphasize edges along an axis where delta is calculated in different resolutions depending on the delta width setting.

In such a case, an alternative input data organization method would be more beneficial. We fed the delta features into ConvNet as with other samples in the same class. By doing this, we expect that the very early stage (i.e., the first layer) of the ConvNet (which usually provides functionality that is similar to edge detection) benefits from delta features by making full use of edge-emphasized versions. This is because ConvNet now has more examples from which to learn about the local characteristics.

To emphasize the spectral characteristics of the audio signal in various resolutions, delta features were extracted from the frequency domain of spectrograms with several different widths. The delta features were calculated as follows:

\begin{equation}
d_f = \frac{\sum_{k=1}^{K}k(x_{f+k}-x_{f-k})}{2\sum_{k=1}^{2}k^2}
\end{equation}

\noindent where $d_f$ represents the delta features in frequency bin $f$, and $K$ represents the number of previous and next frequency bins in the delta calculation, following the conventional method of calculating delta features in the time domain. The term `delta width' as used here refers to $2K + 1$, and is always odd because the window used in the delta feature calculations is symmetrical. We padded the data at the edges with repeated edge values to maintain a constant the size of the data matrix.

\begin{figure*}[!t]
  \centering
  \includegraphics[keepaspectratio, width=0.9\textwidth]{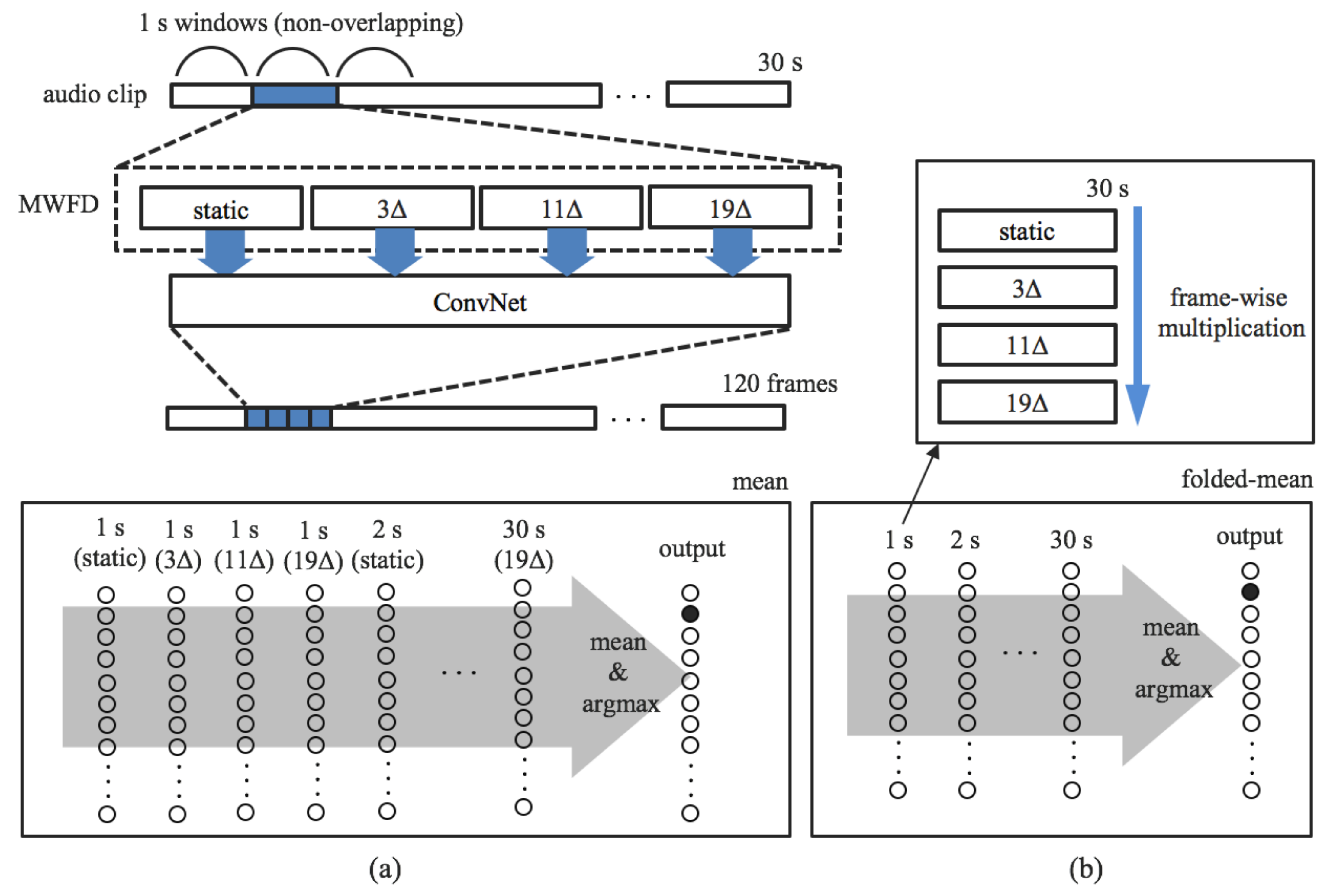}
  \caption{Overall system architecture for MWFD data augmentation, with two different methods for ConvNet output aggregation. (a) All outputs of the static and MWFD data are used as individual outputs, and 120 softmax outputs are averaged over 30 s to enable audio clip-wise decisions. We term this overall mean aggregation (\textit{S1}). (b) Aggregation of static and MWFD data from the same input audio window, first by frame-wise softmax multiplication, and then averaging 30 softmax outputs over 30 s to enable audio clip-wise decisions. We term this folded mean aggregation (\textit{S2}).}
\label{fig:figure_testing}
\end{figure*}

This input data organization method for the delta features can be viewed as data augmentation because it increases the number of input samples; we term this multiple-width frequency-delta (MWFD) data augmentation. For example, consider an input data containing four feature maps is split into four individual examples, each containing a single feature map, as shown in Fig. \ref{fig:figure_spread}. Here the labels for the delta features were copied from the original static input.

\subsection{Aggregating ConvNet Outputs}
For training, 1-s non-overlapping windows were used during the testing phase, but the individual audio clips were 30-s-long. In this section, we describe how we aggregated output probabilities to make an audio clip-wise decision using two different strategies.

With static input, it is relatively straightforward to aggregate output data because each analysis window produces a single output. In this case, taking the mean probability for each class over the audio-clip would be the most straightforward method for aggregation. However, with MWFD, each analysis window produces multiple outputs (four in this case). In such a case, aggregating output probabilities produced from the same audio chunk first to make analysis window-wise decision would be another sensible approach. To this end, we propose a folded mean aggregation which multiplies output probabilities of static and MWFD features from the same window prior to audio clip-wise aggregation process. By multiplying probabilities obtained from different versions of the data, it is possible to maintain classes with high probabilities and suppress the classes where some of the static or MWFD data do not agree.

We illustrate the two methods for softmax output aggregation from ConvNet with the overall system architecture shown in Fig. \ref{fig:figure_testing}. Let aggregation strategy (a) be overall mean aggregation (Fig. \ref{fig:figure_testing}; \textit{S1}) and aggregation strategy (b) be the folded mean aggregation method (\textit{S2}).

\section{Evaluation}
\subsection{Dataset Specifications}
The DCASE 2016 ASC task is essentially an extension of the 2013 task. Several tasks were open as part of the challenge, including ASC, sound event detection in synthetic/real-life audio, and domestic audio tagging; however, in this paper we focus on the ASC task only. Compared with DCASE 2013, the number of scenes for classification had increased from 10 to 15, and the number of audio segments for each scene increased from 10 to 78.

The database for the ASC task contained audio collected in Finland, including scenes of a bus, a cafe/restaurant, a car, a city center, a forest path, a grocery store, a home, a lakeside beach, a library, a metro station, an office, a residential area, a train, a tram, and an urban park. For each class, 78 audio segments were provided for development, as well as a further 26 segments, which were used for evaluation. The audio datasets were recorded using Soundman OKM II Klassik/studio A3, with an electret binaural microphone\footnote{http://www.soundman.de/en/products/} and a Roland Edirol R-09 wave recorder\footnote{http://www.rolandus.com/products/r-09/}. The sampling rate was 44,100 Hz, and the signal was recorded with 24-bit resolution.

\subsection{Baseline System}
The baseline system was provided by the organizer, whereby the audio signals were analyzed with a 40-ms window size and a 20-ms hop size. It used 20 MFCCs, including the zeroth coefficient, along with delta and MFCC acceleration, giving a total of 60 dimensions. GMMs were used as a classifier, with 16 Gaussians to enable scene-level decision making. The system achieved a mean accuracy of 0.725 using an four-fold cross-validation index provided by the organizer, which strictly divides training and testing data such that these do not originate from the same audio source. Further details on these datasets and baseline system can be found in \cite{Mesaros2016_EUSIPCO}.

\subsection{Training Configuration}
We performed network training by optimizing categorical cross-entropy. The network weights were initialized with Glorot uniform data \cite{glorot2010understanding}, and stochastic gradient descent (SGD) with Nesterov momentum \cite{nesterov2007gradient} was used as an optimizer, with a learning rate of 0.02. We decremented the learning rate by 0.0001 for each epoch and the mini-batch size was 128. Randomly selected fifteen percent of the training data were used for validation. To prevent the network from over-fitting the training data, we terminated training when the validation loss stabilized; i.e., whereby the validation loss exhibited no further decrease during 15 epochs. We trained the network using a machine with an NVIDIA GTX 970 GPU and 4 GB of random access memory. The time required for each training epoch was 120 s, giving approximately 2--3 h for each fold.

With reference to Table \ref{tab:convnet_arch}, our network was regularized using dropout, which is widely used for regularization, and randomly removes some units from the network during the training phase only \cite{srivastava2014dropout}. Each of the convolution blocks was regularized with a dropout factor of 0.25. The final fully connected layer had a dropout factor of 0.5.

We performed four-fold cross-validation with the index provided by the organizer. As discussed above, this was identical to the baseline system, and the audio data in each fold were extracted from different recording sources. All experiments were based on the same cross-validation setting, which enabled a performance comparison. The 30-s audio segments were divided into 1-s samples; this was found to be an optimal size for analysis based on empirical experiments, and there were no overlaps between windows. The lengths of each fold differed slightly; each contained between 290 and 297 individual 30-s audio clips. The total number of audio clips was 1168, which provides approximately 26k training examples and 8.7k testing examples. When using MWFD data augmentation, the number of examples was roughly 100k for training and 35k for testing.

\subsection{Experiment Settings}
First, we compared our ConvNet system with the baseline system described above using a `plain' version of ConvNet. In addition, we compared the performance of a deep neural network (DNN) with that of our proposed ConvNet system. With DNN, we used settings that were as similar as possible to that for ConvNet; i.e., using eight layers, with the same window size, hop size, and batch size, as well as the same learning rate and the same optimizer. As DNN requires one-dimensional (1D) data (in contrast to ConvNet), we flattened the 43$\times$128 2D input mel-spec matrices into 5,504 one-dimensional (1D) arrays and fed this to the network. We used normal ReLU for DNN instead of leaky ReLU because it tends to explode the training and validation loss. With DNN, the number of units for each layer was 1024, and classified with 15 softmax. The performance of ConvNet considered here was without the proposed data augmentation and aggregation methods. This was to enable a comparison of the performance of the networks. 

To compare the performance of the input arrangement methods, we carried out an experiment with three different ConvNet input settings. We used the same proposed network architecture and carried out experiments 1) using the original (i.e., static) mel-spectrogram for each example, 2) using MWFD as a 4 feature maps for each example, and 3) with MWFD data augmentation (i.e., the proposed method). Experiments were also performed to compare two different aggregation methods (i.e., \textit{S1} and \textit{S2}) for MWFD data augmentation. The evaluation metric was the accuracy.

\section{Results}
In this section, we discuss the performance of the proposed ConvNet system and compare this with existing algorithms. In addition, we illustrate the effects of the input data augmentation and arrangement method (i.e., MWFD), as well as the influence of the proposed audio clip-wise aggregation method. The results for each setting are the mean accuracy averaged over three repeated experiments for each fold. Different pseudorandom seeding was used for choosing validation data in the repeated experiments; the seeds were fixed across the algorithms to make the comparison as fair as possible.

\begin{table}[!t]
\renewcommand{\arraystretch}{1.3}
\caption{ASC performance of the baseline system (MFCCs + GMM), DNN, and the proposed ConvNet system. The MFCCs of the baseline system include delta and double delta; here the performance metrics for ConvNet are without the proposed input data arrangement and aggregation strategies.}
\centering
\begin{tabular}{l | c}
    \hline
    \hline
    Algorithms & Mean Accuracy\\
    \hline
    MFCCs + GMM & 0.725\\
    DNN & 0.728\\
    ConvNet & 0.778\\
    \hline
    \hline
\end {tabular}
\label{tab:exp_result_algo}
\end{table}

\subsection{Comparison of Algorithms}
As shown in Table \ref{tab:exp_result_algo}, the proposed ConvNet architecture outperformed the baseline system. DNN also exhibited slightly better performance than the baseline system; however, the mean difference of only 0.003 does not represent a significant improvement. These results indicate that exploiting spatially local correlations in the time--frequency representation of the audio signal (as in the ConvNet architecture) is suitable for ASC. DNN does not use local correlation and did not show significant performance improvement over the baseline system, even though it does learn features using a deep neural network. The localized characteristics of ConvNet can be seen as a combination of spectral and temporal information for events that exist in acoustic scenes.

\begin{table}[!t]
\renewcommand{\arraystretch}{1.3}
\caption{ASC performance with the proposed ConvNet system with various input data arrangement methods, aggregation strategies, and ensemble models. Note that MWFD with a four-channel feature map was aggregated using \textit{S1} only. This is because \textit{S2} uses individual output probabilities from deltas, as with the proposed MWFD spread.}
\centering
\begin{tabular}{l | C{1.8cm} | C{1.8cm}}
    \hline
    \hline
    Input data & Mean Acc. & Ensemble Acc.\\
    \hline
    Static mel-spectrogram & 0.778 & 0.786\\
    MWFD (4ch feature map, \textit{S1}) & 0.761 & 0.784\\
    MWFD (spread, \textit{S1}) & 0.814 & 0.820\\
    MWFD (spread, \textit{S2}) & 0.820 & 0.831\\
    \hline
    \hline
\end {tabular}
\label{tab:exp_result_inputdata}
\end{table}

\subsection{Comparison of Input Data Arrangement Methods}
We compared three different ConvNet input scenarios. First, we demonstrate the performance with a static mel-spectrogram as the ConvNet input, which is the most basic approach. Second, we use MWFD features as a four-channel feature map as with color map input in computer vision tasks. Finally, we use MWFD augmentation with the proposed method; i.e., feeding each MWFD data as individual examples. We used three widths for the MWFD: 3, 12 and 19, which covers a wide range of frequencies in the mel-spectrogram with 128 bins. 

As shown in Table \ref{tab:exp_result_inputdata}, the proposed input data augmentation method (i.e., MWFD spread) significantly improved the classification performance compared with using the static mel-spectrogram only. However, feeding MWFD features as a four-channel feature map did not provide good performance; the classification accuracy was lower than that with static input only. This result indicates that training ConvNet together with MWFD features rather than using original static input only clearly helps the network to learn a better feature representation as expected by providing edge-emphasized versions with various resolutions, while feature map approach is more suitable when each feature map contains decomposed parts of the original data such as left/right channel of the stereo audio, or color maps of the image.

\subsection{Comparison of Aggregation Methods}
We carried out experiments with two different aggregation methods to identify 30-s audio clips using 1-s non-overlapping windows in ConvNet. As discussed in the aggregating ConvNet outputs section, \textit{S1} averages output probabilities in a class-wise manner to identify the scene. In this case, the number of softmax outputs used for averaging was 120 for MWFD data augmentation, because it increases the number of input data; each 1-s window produces four softmax outputs. Similarly, the proposed \textit{S2} aggregation strategy averages over ConvNet softmax outputs; however, the softmax outputs from the same audio clip are multiplied together first. Hence, 30 outputs were used for averaging.

As shown in Table \ref{tab:exp_result_inputdata}, it was possible to improve performance using \textit{S2}. In the case of using ensemble model, which is described in more detail in the following section, the use of \textit{S2} improved the performance in a bigger margin by achieving an overall accuracy of 0.831. This result demonstrates that although MWFD features were used as individual examples in the network training step, combining static and MWFD features from the same window first in the aggregation process to make a local (i.e., window-scale) decision prior to a global (i.e.,audio clip-scale) decision certainly helps to obtain a robust identification result, because class-wise multiplication step suppresses the noise.

\subsection{Ensemble Model}
It has been reported that combinations of several different predictors can improve performance. This is because results generated using even the same network may slightly differ, and a model ensemble can generalize this problem \cite{krogh1995neural}. Hence, we combined the networks by taking an average over output probabilities, which is one of the most widely used methods for model ensembles. As shown in Table \ref{tab:exp_result_inputdata}, it was possible to obtain performance improvements via the use of a model ensemble, regardless of the input data arrangement method and aggregation strategy.

\subsection{Class-wise Identification Performance}
To analyze the effects of using ConvNet and MWFD augmentation, we compared the class-wise ASC performance of the baseline system, ConvNet with static input only, and Convnet with MWFD input data augmentation. As shown in Table \ref{tab:exp_result_compare}, using ConvNet generally resulted in improved accuracy compared with the baseline system. In particular, the classification accuracy of the library scene increased from 0.504 to 0.847. Although the mean accuracy was marginally improved by using ConvNet, scenes such as the `cafe/restaurant', `city center', `home', `office', and `residential area' exhibited worse accuracy than the baseline.

Using the proposed MWFD data augmentation improved markedly the accuracy for most of the scenes. With MWFD inputs, ConvNet exhibited poorer accuracy in the `cafe/restaurant' scene only compared to the baseline. It showed slightly lower accuracy for the `home' and `office' scenes, but it was almost negligible. It is interesting to note that using MWFD augmentation rather than static input resulted in a slight decrease in accuracy for the `cafe/restaurant' and `grocery store' scenes, whereas all other acoustic scenes benefited from MWFD augmentation. By observing the confusion matrix, we found that the performance drop of MWFD for the `cafe/restaurant' scene could be attributed mainly to confusion with the `grocery store', and for `grocery store' this was due to confusion with the `metro station'. This result demonstrates that MWFD augmentation was beneficial in most cases, but could result in confusion when the input audio contained a large human vocal component. 

\begin{table}[!t]
\renewcommand{\arraystretch}{1.3}
\caption{Class-wise ASC performance of the baseline system, ConvNet with static input, and ConvNet with the proposed MWFD input data augmentation method. ConvNet results were generated with four-fold mean accuracy of ensemble models.}
\centering
\begin{tabular}{l | C{1.5cm} | C{1.5cm} | C{1.5cm}}
    \hline
    \hline
    Acoustic Scene & Baseline & ConvNet\newline(Static) & ConvNet\newline(MWFD)\\
    \hline
    Beach & 0.693 & 0.763 & 0.868\\
    Bus & 0.796 & 0.809 & 0.875\\
    Cafe/Restaurant & 0.832 & 0.784 & 0.731\\
    Car & 0.872 & 0.935 & 0.935\\ 
    City center & 0.855 & 0.801 & 0.898\\ 
    Forest path & 0.810 & 0.952 & 0.988\\
    Grocery store & 0.650 & 0.847 & 0.808\\
    Home & 0.821 & 0.781 & 0.806\\
    Library & 0.504 & 0.847 & 0.897\\
    Metro station & 0.947 & 1.000 & 1.000\\
    Office & 0.986 & 0.879 & 0.958\\
    Park & 0.139 & 0.264 & 0.328\\
    Residential area & 0.777 & 0.740 & 0.838\\
    Train & 0.336 & 0.505 & 0.662\\
    Tram & 0.854 & 0.881 & 0.898\\
    \hline
    \hline
\end {tabular}
\label{tab:exp_result_compare}
\end{table}

\begin{figure}[!t]
  \centering
  \includegraphics[keepaspectratio, width=3.5in]{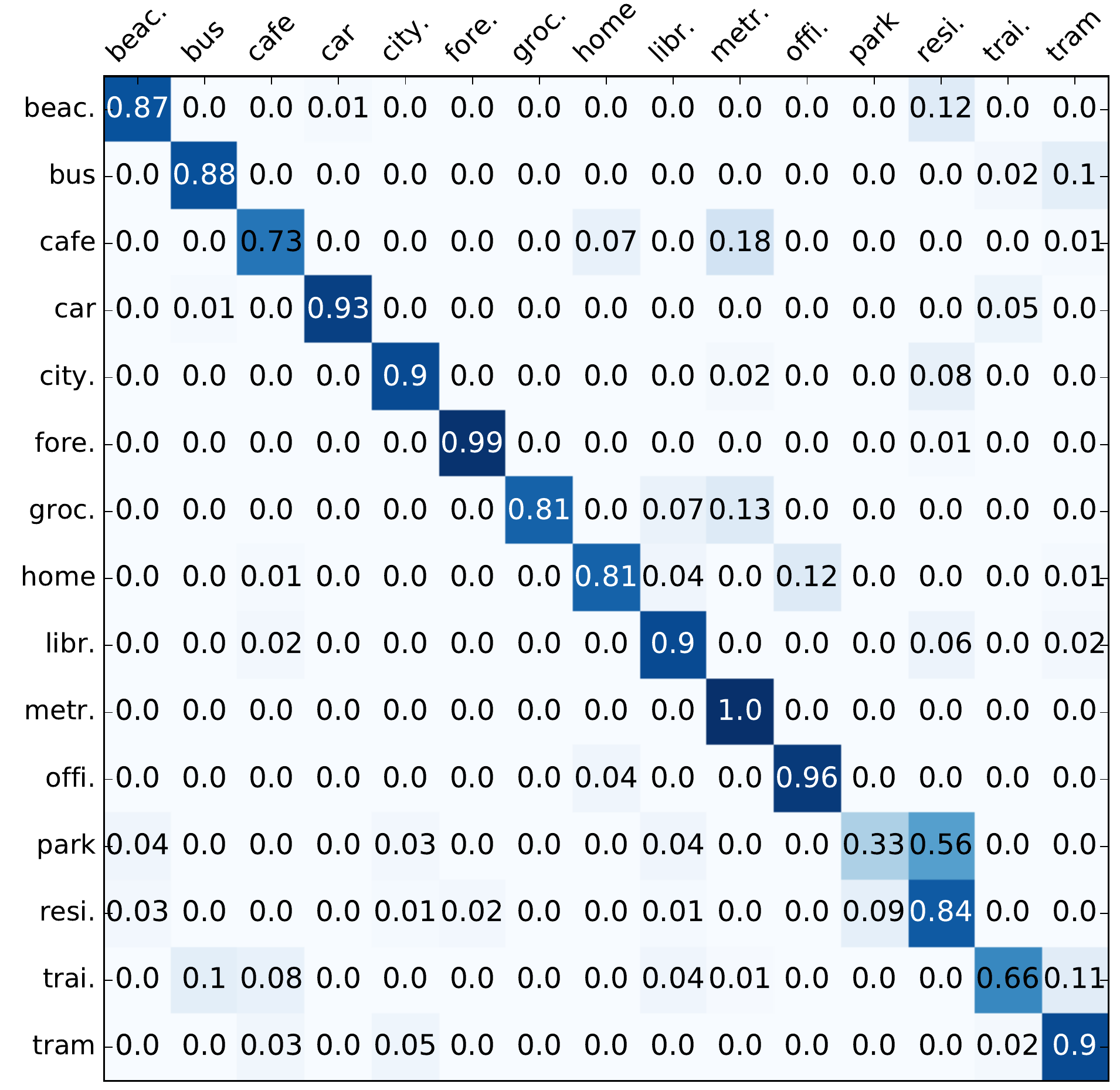}
  \caption{Confusion matrix of the proposed ConvNet system with MWFD data augmentation, extracted from the four-fold cross-validation mean accuracy of the ensemble model with aggregation strategy \textit{S2}, which achieved the best classification result. The scene labels are abbreviated. The original names of the scenes are beach, bus, cafe/restaurant, car, city center, forest path, grocery store, home, library, metro station, office, park, residential area, train, and tram. The $x$-axis is the predicted label; the $y$-axis is the true label.}
\label{fig:figure_conf}
\end{figure}

Among the 15 scenes, 'park' was the most difficult scene to identify, with an accuracy of 0.328, which was much lower than that of the other scenes. From the confusion matrix shown in Fig. \ref{fig:figure_conf}, `park' was often confused with `residential area' -- another relatively quiet outdoor scene. However, it is interesting to note that this confusion was uni-directional; i.e., the `residential area' scene was not commonly confused with the `park' scene, and indeed exhibited a higher than average accuracy. This appears to show that the proposed model could not extract sufficient features from the acoustic signal from the `park' scene, but that the audio signals from the `residential area' scene contained stronger clues enabling differentiation from the other scenes, in addition to `quietness'. By listening to the actual audio clips that formed the training data\footnote{Audio clips available at http://www.cs.tut.fi/sgn/arg/dcase2016/task-acoustic-scene-classification}, the `residential area' could be characterized by traffic sounds as well as bird song, whereas the `park' scene included mainly bird song only. From this inspection, we may surmise that traffic noises provided strong clues in the `residential area' scene, and that the `park' scene was confused with `residential area' due to the bird song, especially since the length of the analysis window was 1-s, meaning it could have included only bird song for both classes in some cases.

The ASC performance of the proposed system for some scenes, such as `forest path' and `metro station', was close to perfect, with mean accuracies 0.988 and 1.000, respectively. This is likely due to the particular sound events that these scenes contained. The `forest path' scene was a wide outdoor space, and the audio always included stepping sounds. Similarly, the `metro station' scene included the sounds of the trains arriving and leaving, which is distinctive and, hence, straightforward to characterize.

\subsection{DCASE 2013 Database Experiment}
\begin{figure}[t]
  \centering
  \includegraphics[keepaspectratio, width=3.5in]{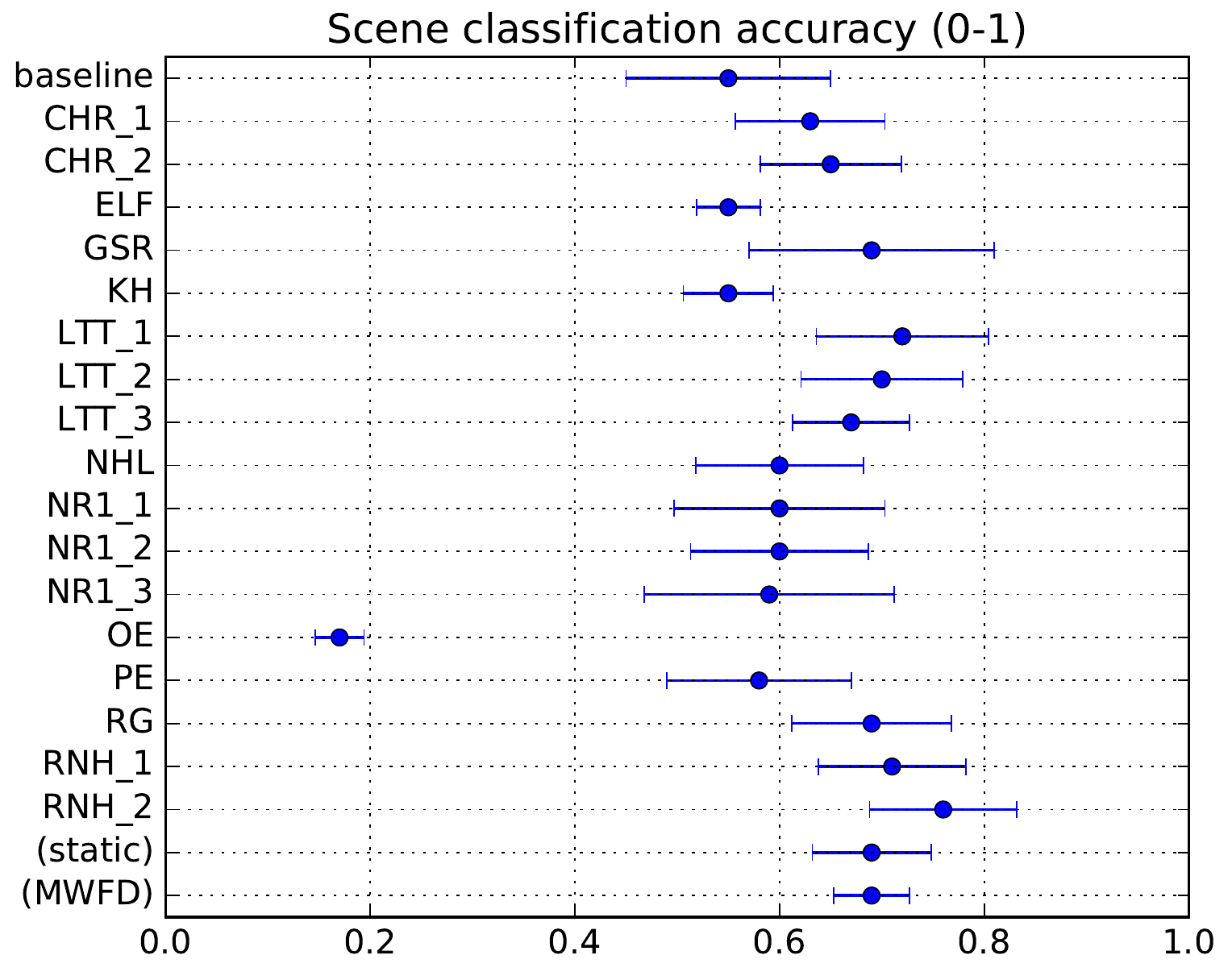}
  \caption{Mean and standard deviation of the five-fold cross-validation accuracy for the DCASE 2013 dataset using the proposed method and other algorithms. Results using the proposed ConvNet approach are shown in brackets (one set used static input, the other used MWFD-augmented input).}
\label{fig:figure_2013}
\end{figure}

We carried out an experiment on the DCASE 2013 database using the proposed ConvNet algorithm to compare the performance of our method with that of existing algorithms. The DCASE 2013 dataset contained 10 acoustic scenes: a bus, a busy street, an office, an open-air market, a park, a quiet street, a restaurant, a supermarket, a tube, and a tube station. The classification results of the other algorithms are taken from a paper summarizing the DCASE 2013 challenge \cite{giannoulis2013detection}. We included results for the proposed ConvNet method with both static-only and MWFD data augmentation in Fig. \ref{fig:figure_2013}. For this experiment, we used five-fold cross-validation on the DCASE 2013 private dataset\footnote{http://c4dm.eecs.qmul.ac.uk/sceneseventschallenge/description.html} to make the experimental conditions identical to that of other algorithms. Details of the other algorithms can be found in Ref. \cite{giannoulis2013detection}.

\begin{figure*}[!t]
  \centering
  \includegraphics[keepaspectratio, width=\textwidth]{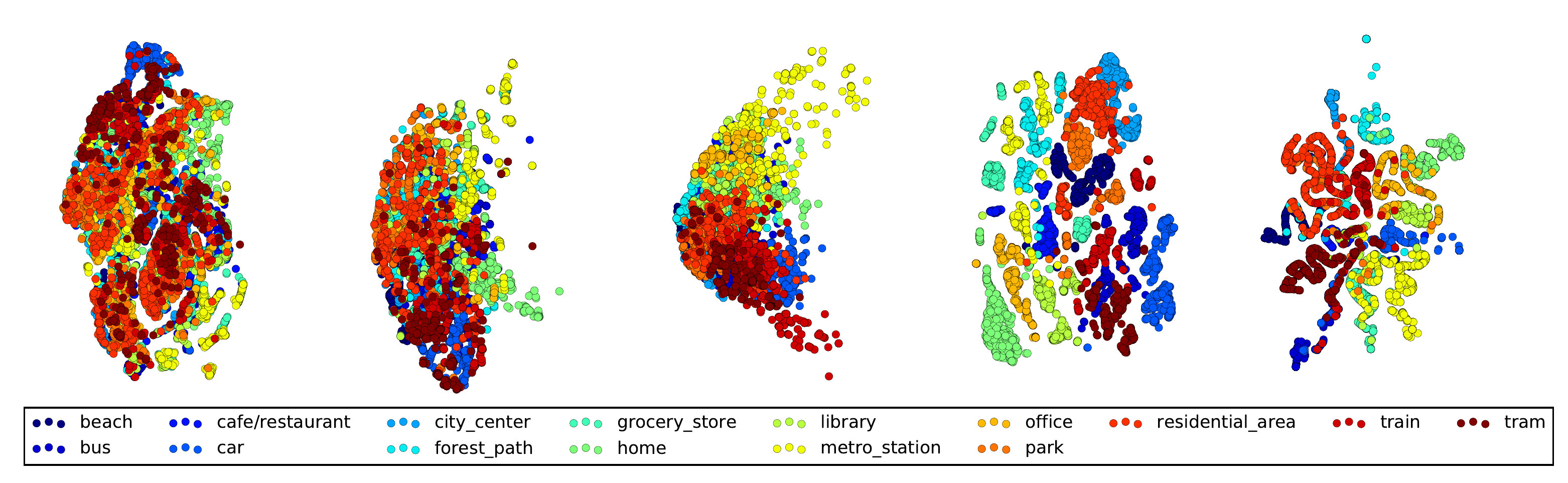}
  \caption{An example of the 2D projection of the intermediate activation of the proposed model using the t-SNE algorithm. Here we show the fold that was not used for training from the four-fold cross-validation, and randomly selected 20\% of data is used for visualization. From the left, the first to fourth plots show the intermediate activations at the end of each convolution block, and the fifth plot shows the final softmax layer output. This reveals that the deep architecture of the model successfully distinguished the data and it gradually became more separable as they go through each of the convolution blocks.}
\label{fig:figure_tsne}
\end{figure*}

As shown in the figure Fig. \ref{fig:figure_2013}, the proposed ConvNet system achieved an accuracy of 0.69, both with static and MWFD augmented input data. This matched or outperformed most other ASC algorithms; however, the accuracy was not the best among submitted algorithms for DCASE 2013. In addition, MWFD data augmentation did not improve the classification accuracy compared with static input.

The main reason for these results is most likely lack of training data. The number of training samples is a critical parameter affecting learning and, hence, good feature representation, especially for a deep neural network. The DCASE 2013 dataset contained 10 segments per scene, and was significantly smaller than the DCASE 2016 dataset, which included 78 segments per scene. There were only eight segments for training and two segments for testing for each scene in the five-fold cross-validation experiment. Except for the algorithm of Nam et al. \cite{nam2013acoustic} which used sparse RBMs for feature learning (see `NHL' in Fig. \ref{fig:figure_2013}), all other algorithms submitted to DCASE 2013 used manual features, which suffers much less from the small quantity of data compared with feature learning methods.

This result suggests that deep learning approaches such as proposed ConvNet system require large training datasets to achieve good accuracy. Furthermore, MWFD data augmentation can be effective when there is a sufficiently large quantity of input data, as it exhibited identical performance to the static input, with only a small decrease in the standard deviation of the accuracy.

\subsection{Qualitative Analysis using t-SNE visualization}
We carried out a visual analysis using t-distributed stochastic neighbor embedding (t-SNE). This minimizes the Kuller--Leibler (KL) divergence between the original feature space and the low-dimensional embedding, and has been reported to be effective for dimensional reduction of data that has very high dimensionality \cite{van2008visualizing}. It is widely used for visualization in various fields \cite{2016arXiv160509507H,platzer2013visualization,van2010texton}. 

To visualize the feature-learning process step by step, we extracted intermediate outputs from each convolution block, as well as the final softmax outputs. Due to the large size of the datasets, we randomly selected 20\% of the data from the test set of one of the cross-validation settings, and pooled the maximum values from each activation matrix. Although the test labels were not used for training, we colored the plots with the ground truth label for visual inspection purposes. Fig. \ref{fig:figure_tsne} shows the t-SNE visualization results, and it shows that the input data gradually became more separable as they through each of the convolution blocks. It is difficult to see any clues from the first two convolution blocks, but the fourth convolution block shows that the scenes became grouped reasonably well.

\section{Conclusion}
We have described an approach to applying ConvNet to the ASC task of the DCASE 2016 challenge. The proposed ConvNet system was composed of eight convolution layers with a leaky ReLU activation function, and max-pooling layers at the end of each pair of convolution layers. Using this ConvNet approach, we achieved a mean classification accuracy of 0.778, which is greater than that using DNN, and was superior to the DCASE 2016 baseline system (which used MFFCs and GMMs).

To improve further the classification accuracy, we implemented MWFD data augmentation, which uses delta with various widths in the frequency domain of the mel-spectrogram and put together with the static input data in the network as individual examples with identical labels. In addition, we demonstrated an effective method to aggregate probabilities from individual analysis windows to enable audio clip-wise decision making, which we term folded mean aggregation. Using the proposed MWFD data augmentation approach with the folded mean aggregation method, we achieved a mean accuracy of 0.820, which represents a significant improvement compared with plain ConvNet; an accuracy of 0.831 was achieved using an ensemble model.

The proposed MWFD data augmentation approach with the folded mean aggregation method represents a simple, highly generalized approach, and is not task-specific. We believe that many other audio processing tasks would benefit from using these methods, and we are planning to apply this approach to a range of other applications.

In this work, we did not make use of several recently introduced neural network techniques such as deep residual learning \cite{he2015deep} and batch normalization \cite{ioffe2015batch}, because these techniques did not provide a better classification accuracy in our empirical experiment with their original settings described in the paper. However, we think that these techniques would be also highly helpful to improve the acoustic scene classification performance further once appropriate network architecture modification and parameter tuning are done. We are planning to investigate on how to blend these techniques into our network and use along with the proposed MWFD augmentation method.


%



\section*{Acknowledgment}
This work was supported by Ministry of Science, ICT (Information and Communication Technologies) and Future Planning by the Korean Government (NRF-2015M3A9D7066980).




\bibliographystyle{IEEEtran}
\bibliography{bare_jrnl}
%



%

\begin{IEEEbiography}[{\includegraphics[width=1in,height=1.25in,clip,keepaspectratio]{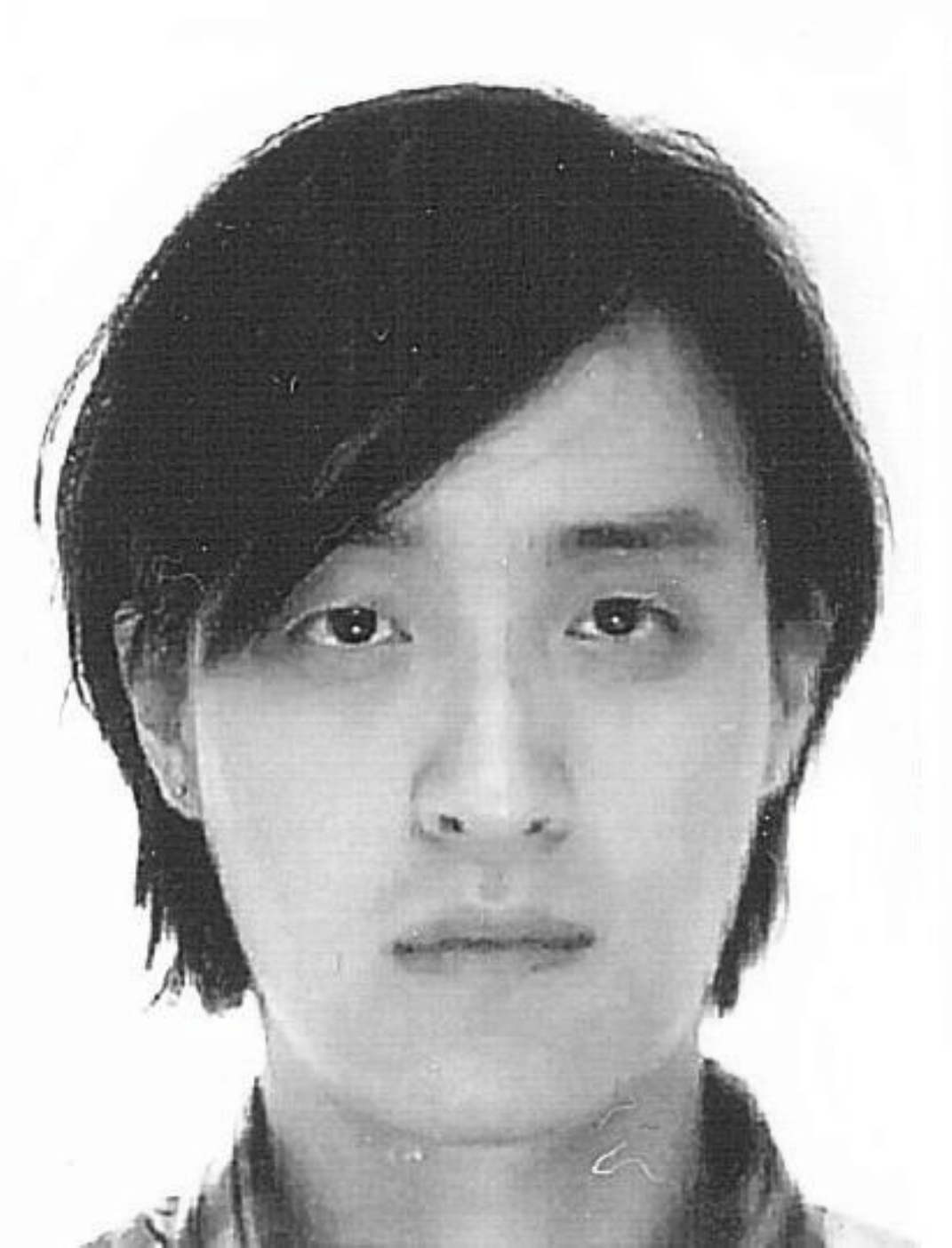}}]{Yoonchang Han}
was born in Seoul, Republic of Korea, in 1986. He studied electronic engineering systems at King's College London, UK, from 2006 to 2009, and then moved to Queen Mary, University of London, UK, and received an MEng (Hons) degree in digital audio and music system engineering with First Class Honours in 2011. He is currently a PhD candidate in digital contents and information studies at the Music and Audio Research Group (MARG), Seoul National University, Republic of Korea. His main research interest lies within developing deep learning techniques for automatic musical instrument recognition.
\end{IEEEbiography}


\begin{IEEEbiography}[{\includegraphics[width=1in,height=1.25in,clip,keepaspectratio]{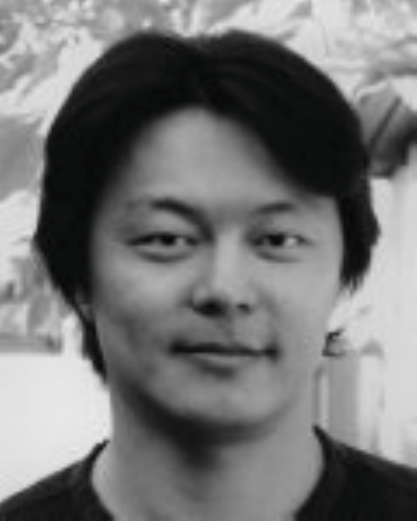}}]{Kyogu Lee}
is an associate professor at Seoul National University and leads the Music and Audio Research Group. His research focuses on signal processing and machine learning techniques applied to music and audio. Lee received a PhD in computer-based music theory and acoustics from Stanford University.
\end{IEEEbiography}


\vfill

\end{document}